\documentstyle[12pt,epsf]{article}

\def\m{\mu}
\def\n{\nu}
\def\r{\rho}
\def\s{\sigma}
\def\p{\phi}
\def\vp{\varphi}
\def\th{\theta}
\def\b{\beta}
\def\a{\alpha}
\def\l{\lambda}
\def\pa{\partial}
\def\ep{\epsilon}

\def\be{\begin{eqnarray}}
\def\ee{\end{eqnarray}}
\def\nn{\nonumber}

\def\ll{\left}
\def\rr{\right}
\def\fr{\frac}

\def\ci{\cite}
\def\bi{\bibitem}

\begin{document}

\title
{\bf \LARGE Gravitating monopoles and black holes 
in Einstein-Born-Infeld-Higgs model}
\author{Prasanta K. Tripathy\thanks{E-mail: prasanta@iopb.res.in}\\
{\small\sl Institute of Physics, Bhubaneswar 751 005, India}}
\maketitle

\begin{abstract}
We find static spherically symmetric monopoles in 
Einstein-Born-Infeld-Higgs model in $3+1$ dimensions. The solutions 
exist only when a parameter $\a $ (related to the strength of Gravitational
interaction) does not exceed certain critical value. We also 
discuss magnetically charged non Abelian black holes in this model. 
We analyse these solutions numerically.
\end{abstract}

\section{Introduction}

Some time ago monopoles in Einstein-Yang-Mills-Higgs(EYMH) model
\ci{{peter},{ortiz},{dieter}}, for $SU(2)$ gauge group with Higgs field in 
adjoint representation, were studied as a generalization of the 
't Hooft-Ployakov monopole\ci{thooft} to see the effect
of gravity on it. In particular, it was found that solutions exist up to 
some critical value of a dimensionless parameter $\a $, characterising the 
strength of the gravitational interaction, above which there is no regular 
solution. The existance of these solutions were also proved analytically 
for the case of infinite Higgs mass\ci{dieter}. Also, non Abelian magnetically 
charged black hole solutions were shown to exist in this model for both 
finite\ci{lee} as well as infinite\ci{bizon} value of the coupling constant 
for Higgs field. The Abelian black holes exists for $r_h \ge \a $ and non 
Abelian black holes exist in a limited region of the $(\a , r_h)$ plane. 

Recently Born-Infeld theory \ci{born, infeld} has received wide publicity, 
especially in the context of string theory\ci{string}. 
Bogomol'nyi-Prasad-Sommerfield (BPS) saturated solutions were obtained in 
Abelian Higgs model as well as in $O(3)$ sigma model in $2+1$ dimensions 
in presence of Born-Infeld term\ci{tuku}. Different models for domain wall, 
vortex and monopole solutions, containing the Born-Infeld Lagrangian 
were constructed\ci{gib} in such a way that the self-dual equations are 
identical with the corresponding Yang-Mills-Higgs model. Recently non 
self-dual monopole solutions were found numerically in non Abelian 
Born-Infeld-Higgs theory\ci{schap}.

In this paper we consider the Einstein-Born-Infeld-Higgs(EBIH) model and 
study the monopole and black hole solutions. The solutions are qualitatively
similar to those of EYMH model. The black hole configurations have
nonzero non Abelian field strength and hence they are 
called non Abelian black holes\ci{pbizon}. In Sec. II we consider the model
and find the equations of motion for static spherically symmetric fields.
In Sec III we find the asymptotic behaviours and discuss the numerical
results. Finally we conclude the results in Sec. IV.

\section{The Model}

We consider the following Einstein-Born-Infeld-Higgs action for $SU(2)$ 
fields with the Higgs field in the adjoint representation 

\be
S = \int d^4x \sqrt{-g} \ll[L_G + L_{BI} + L_H \rr]
\ee
with
\be
L_G & = & \fr{1}{16\pi G}{\cal R} , \nn \\
L_H & = & -\fr{1}{2} D_{\m}\p ^a D^{\m}\p ^a
          -\fr{e^2g^2}{4}\ll(\p ^a\p ^a - v^2 \rr)^2 \nn
\ee
and the non Abelian Born-Infeld Lagrangian\ci{nonab},
\be
L_{BI} = \b ^2 Str\ll( 1 - \sqrt{ 1 
+ \fr{1}{2\b ^2}F_{\m\n}F^{\m\n}
- \fr{1}{8\b ^4}\ll(F_{\m\n}\tilde{F}^{\m\n}\rr)^2}\rr) \nn
\ee
where 
\be
D_{\m}\p ^a = \pa _{\m}\p ^a + e \ep ^{abc} A_{\m}^b\p ^c , \nn
\ee
\be
F_{\m\n} = F_{\m\n}^a t^a 
= \ll(\pa _{\m}A_{\n}^a - \pa_{\n}A_{\m}^a 
+ e \ep ^{abc}A_{\m}^bA_{\n}^c\rr)t^a \nn
\ee
and the symmetric trace is defined as 
\be
Str\ll(t_1,t_2...,t_n\rr) = \fr{1}{n!}
\sum tr\ll(t_{i_1}t_{i_2}...t_{i_n}\rr) \nn .
\ee
Here the sum is over all permutations on the product of 
the $n$ generators $t_i$. Here we are interested in purely magnetic
configurations, hence we have $F_{\m\n}\tilde{F}^{\m\n} = 0 $. 
Expanding the square root in powers of $\fr{1}{\b ^2}$ and keeping 
up to order  $\fr{1}{\b ^2}$ we have the Born-Infeld Lagrangian 

\be
L_{BI} = -\fr{1}{4}F_{\m\n}^a F^{a \m\n}
+\fr{1}{96\b ^2}\ll[\ll(F_{\m\n}^a F^{a \m\n}\rr)^2
+2F_{\m\n}^a F_{\r\s}^a F^{b \m\n}F^{b \r\s}\rr] 
+ O(\fr{1}{\b ^4}) .  \nn
\ee
For static spherical symmetric solutions, the metric can be 
parametrized as\ci{dieter, komar}

\be
ds^2 = -e ^{2\n(R)}dt^2 + e ^{2\l(R)}dR^2 
+ r^2(R)(d\th ^2 + \sin ^2\th d\vp ^2)
\ee
and we consider the following  ansatz for the gauge  and scalar fields 

\be
A_{t}^a(R) = 0 = A_{R}^a, A_{\th}^a = e_{\vp}^a\fr{W(R) - 1}{e}, 
A_{\vp}^a = -e_{\th}^a\fr{W(R) - 1}{e}\sin\th ,
\ee
and 
\be
\p ^a = e_{R}^a v H(R) .
\ee
Putting the above ansatz in Eq.1, defining $\a ^2 = 4\pi Gv^2$ 
and rescaling $ R \rightarrow R/ev, \b \rightarrow \b ev^2 $ and
$ r(R) \rightarrow r(R)/ev $ we get the following expression for 
the Lagrangian
\be
\int dR e^{\n +\l}\ll[\fr{1}{2}\ll(1
+ e^{-2\l}\ll((r')^2 + \n '(r^2)'\rr)\rr)
- \a ^2 \ll(e^{-2\l} V_1 - e^{-4\l} V_2 + V_3 \rr)\rr],
\ee
where 
\be
V_1 = (W')^2 +\fr{1}{2}r^2(H')^2 - (W')^2\fr{(W^2 - 1)^2}{6\b ^2r^4},
\ee
\be
V_2 = \fr{(W')^4}{3\b ^2r^2}
\ee 
and 
\be
V_3 = \fr{(W^2 - 1)^2}{2r^2} + W^2H^2 + \fr{g^2r^2}{4}(H^2 - 1)^2 
- \fr{(W^2 - 1)^4}{8\b ^2r^6} . 
\ee
Here the prime denotes differentiation with respect to $R$. 
The dimensionless parameter $\a $ can be expressed as the mass ratio
\be
\a = \sqrt{4\pi}\fr{M_W}{eM_{Pl}}
\ee
with the gauge field mass $M_W = e v $ and the Planck mass 
$M_{Pl} = 1/ \sqrt{G} $ . Note that the Higgs mass $M_H = \sqrt{2} g e v $. 
In the limit of $\b \rightarrow \infty $ the above action reduces to 
that of the Einstein-Yang-Mills-Higgs model\ci{peter,ortiz}. 
For the case of $\a = 0$ we must have $\n (R) = 0 = \l (R)$ which 
corresponds to the flat space Born-Infeld-Higgs theory\ci{schap}.
We now consider the gauge $r(R) = R $, corresponding to the 
Schwarzschild-like coordinates and rename $R = r $. 
We define $A = e^{\n + \l}$ and $N = e^{-2\l}$. 
Varying the matter field Lagrangian with respect to the metric we find 
the energy-momentum tensor. Integrating the $tt$ component of the 
energy-momentum we get the mass of the monopole equal to $M/evG$ where 
\be
M = \a ^2 \int_{0}^{\infty} dr \ll(NV_1 - N^2V_2 + V_3\rr)
\ee

Following 't Hooft the electromagnetic $U(1)$ field strength 
${\cal F}_{\m\n}$ can be defined as
\be
{\cal F}_{\m\n} = \fr{\p ^aF_{\m\n}^a}{\mid \p \mid}
- \fr{1}{e\mid\p\mid ^3}\epsilon^{abc}\p ^aD_{\m}\p ^bD_{\n}\p ^c. \nn
\ee
Then using the ansatz(3)  the magnetic field 
\be B^i = \fr{1}{2}\epsilon ^{ijk}{\cal F}_{jk} \nn \ee is equal to 
$ {e_{r}^{i}}/{er^2} $ with a total flux 
$4\pi /e $ and unit magnetic charge.

The $tt$ and $rr$ components of Einstein's equations are 
\be
&&\fr{1}{2}\ll(1 - (rN)'\rr)  =  \a ^2\ll( N V_1 - N^2 V_2 +V_3 \rr)  \\
&&\fr{A'}{A}  = \fr{2\a ^2}{r}\ll(V_1 - 2NV_2\rr). 
\ee
The equations for the matter fields are 
\be
&&\ll(ANV_4\rr)'  = A W \ll(
\fr{2}{r^2}(W^2 - 1) + 2 H^2 - \fr{(W^2 - 1)^3}{\b ^2 r^6}
-\fr{2N(W')^2}{3\b ^2 r^4} (W^2 - 1) \rr) \\
&&(ANr^2H')'  = A H \ll(2W^2 + g^2r^2(H^2 - 1)\rr)
\ee
with  
\be
V_4 = 2W' - \fr{W'}{3\b ^2 r^4}(W^2 - 1)^2 
- \fr{4N}{3\b ^2 r^2}(W')^3
\ee
It is easy to see that $A$ can be elliminated from the matter  field 
equations using Eq.(12). Hence we have to solve three differential equations 
Eqs. (11),(13) and (14) for the three fields $N, W$ and $H$.
  
\section{Solutions}

\subsection{Monopoles}

For finite $g$, demanding the solutions to be regular and the 
monopole mass to be finite gives the following behaviour near the origin  
\be
&&H = a r + O(r^3), \\
&&W = 1 - b r^2 + O(r^4), \\
&&N = 1 - c r^2 + O(r^4),
\ee
where $a$ and $b$ are free parameters and $c$ is given by
\be
c = \a ^2 \ll( a^2 + 4b^2 + \fr{g^2}{6} - \fr{20 b^4}{3\b ^2} \rr). \nn
\ee
In general, with these initial conditions $N$ can be zero at some finite
$r$ where the solutions become singular. In order to avoid this singularity
we have to adjust the parameters $a$ and $b$ suitably. 

For $r \rightarrow \infty $ we require the solutions to be asymptotically
flat. Hence we impose 
\be
N = 1 - \fr{2M}{r}
\ee
Then for finite mass configuration we have the following expressions for 
the gauge and the Higgs fields
\be
&& W = C r^{-M} e^{-r}\ll(1 + O(\fr{1}{r})\rr) \\
&& H = \left\{ \begin{array}{ll}
1 - B r^{-\sqrt{2}gM - 1} e^{-\sqrt{2}gr}, & for ~~~ 0< g \le \sqrt{2} \\
1 - \fr{C^2}{g^2 - 2} r^{-2M-2} e^{-2r},   & for ~~ g = 0 ~~~ and g > \sqrt{2}.
\end{array}
\right.
\ee
Note that the fields have similar kind of asymptotic behaviour in the 
EYMH model\ci{dieter}.
We have solved the equations of motion numerically with the boundary 
conditions given by Eqs.(16-21). For $\a=0$, $g=0$ and $\b \rightarrow \infty $ 
they corresponds to the exact Prasad-Sommerfield solution\ci{prasad}. 
For nonzero $\a ,g$ and finite $\b $ the qualitative behaviour of the solutions
are similar to the corresponding solutions of EYMH model. For large $r$ these 
solutions converges to their asymptotic values given as in Eqs.(19-21).  
For a fixed value of $g$ and $\b $ we solved the equations increasing the 
value of $\a $. For small value of $\a $ the solutions are very close to 
flat space solution. As $\a $ is increased the minimum of the metric 
function $N$ was found to be decreasing. The solutions cease to exist 
for $\a $ greater then certain critical value $\a _{max}$. 
For $g=0$ and $\b =3$ we find $\a _{max} \sim 2 $. The profile for the 
fields for different values of $\a $ with $g = 0$ and $\b = 3$ are given 
in Figs.1,2 and 3. The profile for the fields for $g=.1, 
\a =1.0 $ and $\b = 3 $ are given in Fig. 4. We find numerically 
the mass $M = 0.7865 $ of the monopole  for $g=.1,\a =1.0$ and $\b = 3$.

\subsection{Black holes}

Apart from the regular monopoles, magnetically charged black holes can 
also exist in this model. Black hole arises when the field $ N $ vanishes
for some finite $r = r_h$ . Demanding the solutions to be regular near 
horizon $r_h$ we find the following behaviour of the fields 
\be
&& N(r_h + \r ) = N_{h}'\r + O(\r ^2) ,\\
&& H(r_h + \r ) = H_h + H_{h}'\r + O(\r ^2), \\
&& W(r_h + \r ) = W_h + W_{h}'\r + O(\r ^2) 
\ee
with
\be
&& N_{h}' = \fr{1}{r_h}\ll[ 1 - \a ^2\ll\{
\fr{(W_{h}^2 - 1)^2}{r_{h}^2} + 2 W_{h}^2 H_{h}^2 
+\fr{g^2r_{h}^2}{2}(H_{h}^2 - 1)^2 
-\fr{(W_{h}^2 - 1)^4}{4\b ^2 r_{h}^2} \rr\}\rr] \\
&& H_{h}' = \fr{H_{h}}{N_{h}'r_{h}^2}
\ll\{ 2W_{h}^2 + g^2 r_{h}^2 (H_{h}^2 - 1)\rr\} \\
&& W_{h}' = \fr{
\fr{W_h}{r_{h}^2}(W_{h}^2 - 1) + W_h H_{h}^2
- \fr{W_h}{2\b ^2 r_{h}^6}(W_{h}^2 - 1)^3}
{N_{h}' \ll[ 1 - \fr{(W_{h}^2 - 1)^2}{6\b ^2r_{h}^4}\rr]}.
\ee
Here $r_h, W_h(\equiv W(r_h)) $ and $H_h(\equiv W(r_h))$ are arbitrary. 
For $r \rightarrow \infty $ the behaviour of the fields is same as the 
regular monopole solution as given by Eqs.(19-21). 
The black hole has unit magnetic charge with nontrivial gauge field strength. 
We found numerical solutions to the non Abelian black hole for 
different $r_h$. For a fixed value of $r_h $ we find the solutions for 
$r > r_h $ adjusting the parameters $W_h $ and $H_h $. For $r_h $ close to 
zero the solutions approach the regular monopole solutions. The profile for the 
fields are given in Fig.5. We found the mass of the black hole equals to be 
$ 0.6796 $ for $\a = 1.0, g = 0 $ and $\b = 3$.

\section{Conclusion}

In this paper we have investigated the effect of gravity on the 
Born-Infeld-Higgs monopole. We found that solutions exist only up to some  
critical value $\a _{max}$ of the parameter $\a $. In the limit 
$\b \rightarrow \infty $ these solutions reduces to those of EYMH monopoles. 
We also found numerically magnetically charged non Abelian black hole 
solutions in this model. It would be interesting to prove analytically the 
existence of these solutions for finite value of the parameters. Recently 
dyons and dyonic black holes were found in EYMH model numerically\ci{tell} 
and the existence of critical value for $\a $ was also proved analytically
\ci{tel}. It may be possible to generalize these solutions to find dyons and 
dyonic black holes in EBIH model. We hope to report on this issue in future.

\section{Acknowledgements}

I am indebted to Avinash Khare for many helpful discussions as well as for 
a careful manuscript reading.

\newpage
\begin{figure}
\vspace{-1in}
\vglue.1in
\makebox{
\epsfxsize=9in
\epsfbox{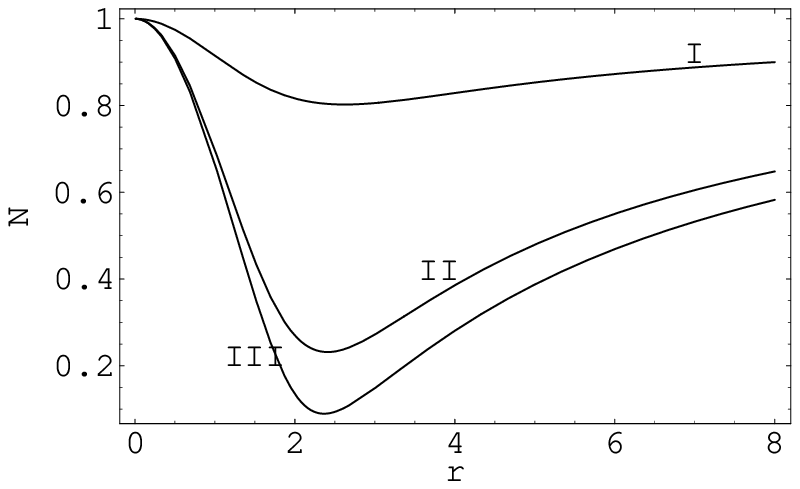}
}
\vspace{-9in}
\caption{Plot for the metric function $N$ as a function of $r$ 
for $g = 0$, $\b = 3 $ for different values of $\a $. Curve I is 
for $\a = .7$, curve II for $\a = 1.6$ and curve III for $\a = 1.9 $. 
}


\vglue.1in
\makebox{
\epsfxsize=9in
\epsfbox{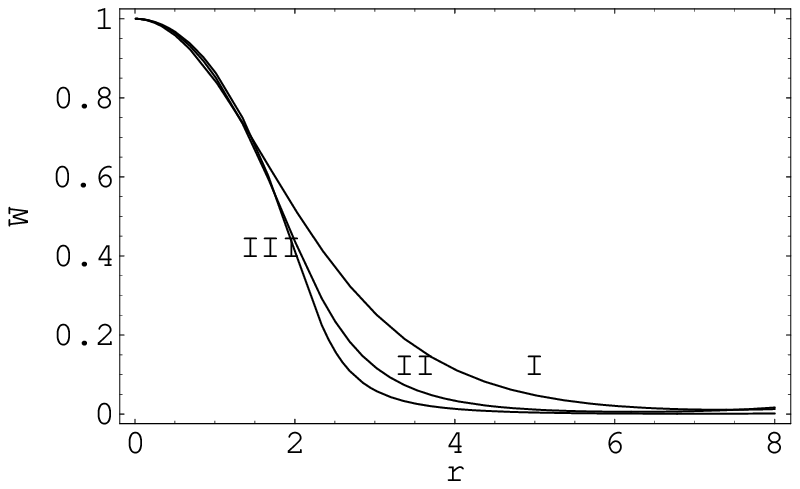}
}
\vspace{-9in}
\caption{Plot for the gauge field $W$ as  a function of $r$
for $g = 0$, $\b = 3 $ for different values of $\a $.  Curve I for $\a = .7$, 
curve II for $\a = 1.6$ and curve III for $\a = 1.9$. }

\end{figure}

\newpage

\begin{figure}
\vspace{-.5cm}
\vglue.1in
\makebox{
\epsfxsize=9in
\epsfbox{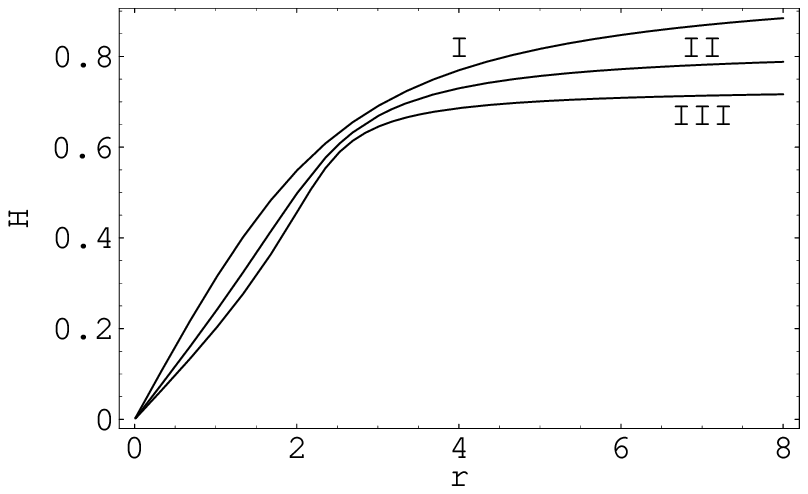}
}
\vspace{-9in}
\caption{Plot for the Higgs field $H$ as a function of $r$ 
for $g = 0$, $\b = 3 $ for different $\a $.  Curve I is for $\a = .7$,
curve II for  $\a = 1.6$ and curve III for $\a = 1.9$. }


\vglue.1in
\makebox{
\epsfxsize=9in
\epsfbox{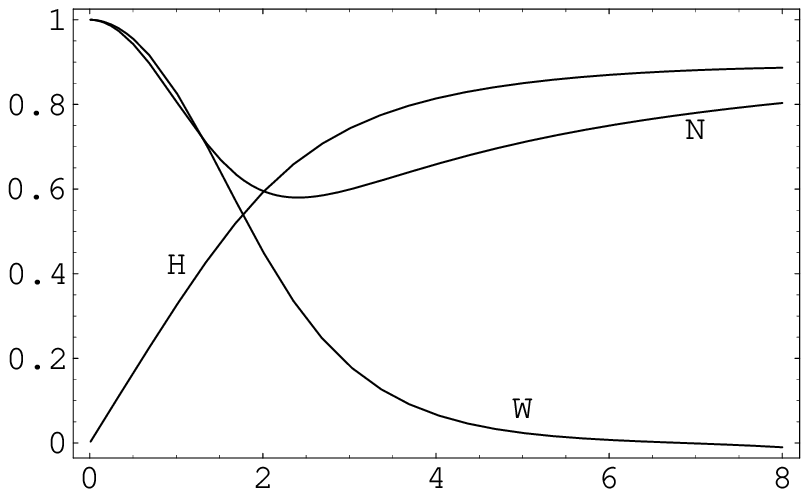}
}
\vspace{-9in}
\caption{Plot for the fields $N$ , $W$ and $H$ as function of $r$ 
for $g = 0.1$ , $\b = 3 $ and $\a = 1 $. }
\end{figure}

\newpage

\begin{figure}
\vspace{-1in}
\vglue.1in
\makebox{
\epsfxsize=9in
\epsfbox{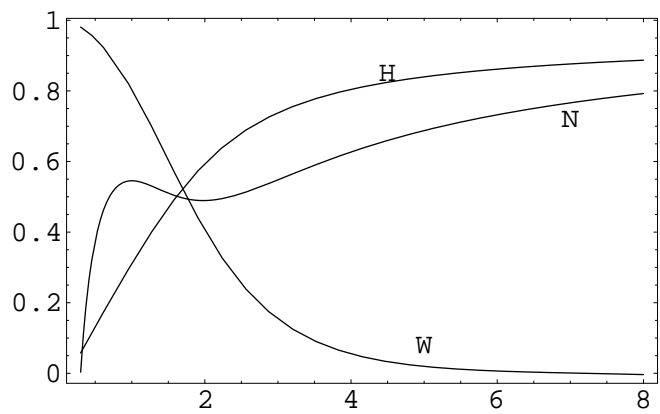}
}
\vspace{-9in}
\caption{Black hole solutions for $g = 0$ , $\b = 3 $, $\a = 1$ , $r_h = 0.3 $,
$H_h = 0.057271 $ and $W_h = 0.980997$.}
\end{figure}

\end{document}